\documentclass[fleqn,usenatbib]{mnras}

\usepackage{newtxtext,newtxmath}
\usepackage[T1]{fontenc}

\DeclareRobustCommand{\VAN}[3]{#2}
\let\VANthebibliography\thebibliography
\def\thebibliography{\DeclareRobustCommand{\VAN}[3]{##3}\VANthebibliography}

\usepackage{graphicx}	% Including figure files
\usepackage{amsmath}	% Advanced maths commands
\usepackage[para,online,flushleft]{threeparttable}

\title[Spectroscopy of G107.7$-$5.1 and G150.3+4.5]{First detailed optical spectroscopic observations of the supernova remnants G107.7$-$5.1 and G150.3+4.5}

\author[E.~Aktekin et al.]{{E.~Aktekin$^{1}$\thanks{{\color {blue}corresponding author: ebrucaliskan@sdu.edu.tr} (EA)},
H.~Bak{\i}\c{s}$^{2}$,
V.~Bak{\i}\c{s}$^{2}$ and
A.~Sezer$^{3}$}\\
$^{1}$Department of Physics, S\"{u}leyman Demirel University, Isparta 32000, T\"{u}rkiye \\
$^{2}$Department of Space Sciences and Technologies, Akdeniz University, Antalya 07058, T\"{u}rkiye\\
$^{3}$Department of Computer Engineering, Avrasya University, Trabzon 61250, T\"{u}rkiye\\
}

\date{Accepted XXX. Received YYY; in original form ZZZ}

\pubyear{\the\year{}}

\begin{document}
\label{firstpage}
\pagerange{\pageref{firstpage}--\pageref{lastpage}}
\maketitle

\begin{abstract}
We present optical spectroscopic observations of the supernova remnants (SNRs) G107.7$-$5.1 and G150.3+4.5, each spanning nearly 3$\degr$. Both remnants were recently examined in the optical band through a deep H$\alpha$ and [\ion{O}{iii}] emission-line imaging survey, which led to the discovery of G107.7$-$5.1. Using long-slit spectra obtained with the 1.5-m Russian-Turkish Telescope (RTT150), we investigate the physical conditions of the pre-shock and post-shock gas in the optical filamentary regions of the SNRs. The SNR nature of G107.7$-$5.1 is confirmed by the measured [\ion{S}{ii}]/H$\alpha$ and [\ion{N}{ii}]/H$\alpha$ ratios, which range from 0.56 to 0.86 and from 0.7 to 1.4, respectively. A similar conclusion is reached for G150.3+4.5, where the observed [\ion{S}{ii}]/H$\alpha$ (0.43$-$0.92) and [\ion{N}{ii}]/H$\alpha$ (0.49$-$1.29) ratios likewise support a shock-excitation origin.  
Further confirmation of the shock-excited nature of both SNRs comes from their consistency with recent diagnostic diagrams based on multiple emission-line ratios. The [\ion{O}{iii}]/H$\beta$ line ratios measured in both remnants indicate shocks with complete recombination zones and are consistent with shock velocities of $\gtrsim$100~km~s$^{-1}$. The electron densities ($n_{\rm e}$), derived from the [\ion{S}{ii}] $\lambda$6716/$\lambda$6731 line ratios, exhibit substantial variation in the spectra of both SNRs. Additionally, extinction variations observed across the remnants suggest the presence of significant dust structures along the line of sight. We conclude that these two remnants display remarkable similarities across multiple diagnostic spectral properties, consistent with previous reports indicating comparable GeV gamma-ray characteristics.
\end{abstract}


%%%%%%%%%%%%%%%%%%%%%%%%%%%%%%%%%%%%%%%%%%%%%%%%%%

%%%%%%%%%%%%%%%%% BODY OF PAPER %%%%%%%%%%%%%%%%%%

\section{Introduction}
%% General
The optical emission from supernova remnants (SNRs) is dominated by strong line radiation, particularly H$\alpha$, along with prominent forbidden transitions of nitrogen, oxygen, and sulphur. Consequently, optical spectroscopy provides crucial diagnostics for probing the physical conditions of SNRs and the gas in their environments (e.g. \citealt{Fesen2008, Fesen2015, Fesen2019, Fesen2020, Fesen2021, Stupar2007, Stupar2009, Stupar2012, Sabin2013, Alikakos2012, Raymond2020, Seok2020, Boumis2022, Palaiologou2022, Aktekin2025}).

The line ratio of [\ion{S}{ii}]/H$\alpha$ is commonly used to distinguish shocked nebulae from photoionized regions, with a canonical threshold of $\geq 0.4$ indicating shock excitation (e.g. \citealt{Fesen1985, Leonidaki2013, Long2017}). Additionally, relatively strong [\ion{N}{ii}]/H$\alpha$ ratios (0.5 $<$ [\ion{N}{ii}]/H$\alpha$ $<$ 1.5) are also indicative of an SNR origin \citep{Fesen2024}. Recent studies have introduced refined diagnostic criteria to distinguish SNRs from photoionized nebulae (e.g. \citealt{Kopsacheili2020, Congiu2023}), extending beyond classical empirical methods by incorporating multiple emission-line ratios. These multiline diagnostics have been widely applied to the identification of extragalactic SNRs and have also been used for Galactic SNRs as an additional test of their shock-excited nature (see HB3: \citealt{Boumis2022} and G116.6$-$26.1: \citealt{Palaiologou2022}).

Two SNRs, G107.7$-$5.1 and G150.3+4.5, each with an angular extent of approximately 3$\degr$, located at relatively high Galactic latitudes and exhibiting an optical filamentary morphology \citep{Fesen2024}, are selected for the present study. Optical spectroscopic analysis of both SNRs will enable a comprehensive investigation of their physical conditions and ionization structures. 

SNR G107.7$-$5.1 (the Nereides nebula) was recently discovered in the optical band by \citet{Fesen2024} using deep H$\alpha$ and [\ion{O}{iii}] emission-line imaging. The authors reported that, while the H$\alpha$ emission appears predominantly diffuse with only a few short, sharp filaments, the [\ion{O}{iii}] emission exhibits an extensive and highly complex filamentary structure. Additionally, \citet{Fesen2024} reported that a short-exposure, low–S/N red spectrum obtained with the MDM 2.4~m telescope from the northwestern limb of the SNR suggests an [\ion{S}{ii}]/H$\alpha$ ratio exceeding 0.4, indicative of shock excitation. The authors emphasized that this measurement requires confirmation with higher-quality data. 

Subsequently, \citet{Araya2024} confirmed the SNR nature of G107.7$-$5.1 through the detection of a counterpart at GeV energies in {\it Fermi}-LAT data. Additionally, the author analysed CO survey data, but did not detect any molecular gas associated with the region of the SNR. \citet{Araya2024} proposed a leptonic origin for the gamma-ray emission and noted that the GeV properties of the SNR are similar to those of other SNRs such as G150.3+4.5. 

\citet{Soker2024} and \citet{Soker2025} identified structural features - including three rims, a nozzle, a possible jet-inflated bubble, and a possible ring - in the [\ion{O}{iii}] line image of G107.7$-$5.1 presented by \citet{Fesen2024}. Based on its highly aspherical morphology, they proposed that it is a core-collapse supernova remnant.

Notably, no radio emission has yet been detected from G107.7$-$5.1. In addition, its exact distance remains uncertain. 

On the other hand, G150.3+4.5 was first considered as an SNR candidate based on the radio detection of its brightest shell region \citep{Gerbrandt2014}. Using the Canadian Galactic Plane Survey (CGPS; \citealt{Taylor2003}), they detected faint radio emission from the southeastern portion of G150.3+4.5's (called G150.8+3.8) shell and identified it as a strong SNR candidate based on its semi-circular shell morphology and non-thermal spectrum. Subsequently, \citet{GaoHan2014} performed radio observations of the SNR and concluded that the observed shell-like structures and their non-thermal emission provide strong evidence that G150.3+4.5 is a shell-type SNR.

Using {\it Fermi}-LAT observations, \citet{Devin2020} examined the morphological and spectral characteristics of G150.3+4.5 and concluded that the gamma-ray emission is explained by a leptonic scenario. Additionally, \citet{Devin2020} analysed {\it ROSAT} data and reported no significant detection of an X-ray emission. 

Using high-resolution optical imaging in H$\alpha$ and [\ion{O}{iii}], \citet{Fesen2024} recently investigated G150.3+4.5 in the optical band and detected numerous [\ion{O}{iii}]-bright filaments across the remnant, particularly along its southern limb. They found excellent positional agreement between the [\ion{O}{iii}] filaments and the radio emission, especially along the southeastern shell, and concluded that the observed optical emission is associated with the SNR.

\citet{Feng2024} examined the molecular environment of G150.3+4.5 and found that the alignment of its morphology, together with the gas kinematics, indicates an association between the SNR and the surrounding molecular clouds (MCs). They suggested that the very high-energy emission detected from 1LHAASO~J0428+5531 \citep{Cao2024} in the direction of the remnant is most likely produced by interactions between the SNR and the nearby MCs, as interpreted by \citet{Zeng2024}.

Using 14~yr {\it Fermi}-LAT data, \citet{Li2024} showed that the spatial correspondence between MCs and gamma-ray emission in the southern region suggests a hadronic origin, whereas the gamma-ray emission from the northern region is likely dominated by relativistic electrons via inverse-Compton processes. They further suggested that this hybrid gamma-ray emission arises from the evolution of the SNR in an inhomogeneous environment.

The distance to SNR G150.3+4.5 remains uncertain. Although gamma-ray analysis suggests that G150.3+4.5 is dynamically young due to the absence of X-ray emission, its distance could only be constrained to 0.7$-$4.7 kpc \citep{Devin2020}. Subsequent studies estimated distances of $\sim$1 kpc based on the hard gamma-ray spectral characteristics of shell-type SNRs \citep{Zeng2021} and $\sim$0.74 kpc from its association with MCs \citep{Feng2024}.

SNRs G107.7$-$5.1 and G150.3+4.5 have not previously been investigated through detailed optical spectroscopy. Consequently, key physical parameters - such as electron density, pre-shock density, and shock velocity - derived from optical spectral diagnostics remain unconstrained. In this study, we present the first optical spectra obtained with the 1.5-m Russian-Turkish Telescope (RTT150)\footnote{\url{https://trgozlemevleri.gov.tr/teleskoplar/antalya/rtt150}} at the T\"{u}rkiye National Observatories\footnote{\url{https://trgozlemevleri.gov.tr/}} and utilize the measured diagnostic line ratios to explore the physical characteristics of both remnants.

Our spectral observations of G107.7$-$5.1 and G150.3+4.5 are described in Section~\ref{sec:obs}, as well as the data reduction. Section~\ref{sec:analysis} presents the measurements of the line fluxes and the physical parameters estimated from the line ratios. In Section~\ref{sec:discuss}, we discuss the implications of the spectral line emission for the nature of the two SNRs and their environments, and Section~\ref{sec:conc} provides a summary of the main results.

%%%%%%%%%%%%%%%%%%%%%%%%%%%%%%%%%%%%%%%%%%%%%%%%%%%%%%%%%%

\section{Spectral observations and data reduction}
\label{sec:obs}
To obtain spectral data, we used the TFOSC (TUG Faint Object Spectrograph and Camera) instrument mounted on the 1.5-m  RTT150 telescope. We used grism 15, which covers the wavelength range 3230$-$9120 {\AA}, providing a spectral resolution of 749. The slit, measuring 2.38 arcsec in width and 234 arcsec in length, was aligned along the east-west direction at all observed positions.

For spectroscopy, we selected positions along the bright H$\alpha$ filaments seen in the upper panel of fig. 31 for G107.7$-$5.1 and in the bottom panel of fig. 19 for G150.3+4.5 of \citet{Fesen2024}. Three exposures of 1200~s were obtained at each slit position, resulting in a total integration time of 3600~s. Table~\ref{tab:Table1} shows the spectroscopic observations log of both SNRs analysed in this work.

Spectral data were reduced and calibrated employing standard techniques in {\sc iraf}\footnote{\url{https://iraf-community.github.io/}} (Image Reduction and Analysis Facility; \citealt{Tody1986, Tody1993}). 1D spectra were extracted using the \texttt{apall} task by defining the appropriate aperture, performing background subtraction, and tracing the spectra. We selected a single extraction aperture for each slit, with an aperture length of approximately 3.3 arcsec used for all slit positions that are free of field stars and contain sufficient line emission to enable accurate measurement of the observed spectral lines. For the background subtraction, regions located east or west of the slit centre were selected, depending on the filament positions, ensuring that they were free of field stars. Wavelength calibration was performed with an iron–argon lamp, and flux calibration was carried out using standard star BD+28 4211 \citep{Oke1990}. 

%%%%%%%%%%%%%%%%%%%%%%%%%%%%%%%%%%%%%%%%%%%%%%%%%%%%%%%%%%
%%%% Table 1: Log of spectroscopic observations
\begin{table*}
\centering
 \caption{The log of spectroscopic observations of G107.7$-$5.1 and G150.3+4.5. The total integration time at each slit position was $3 \times 1200$~s.}
 \label{tab:Table1}
 \begin{threeparttable}
\begin{tabular}{@{}p{2.2cm}p{2.2cm}p{2.2cm}p{2.2cm}p{2.2cm}p{2.4cm}@{}}
 \hline
Slit ID  &    \multicolumn{2}{c}{Slit centres}                 &  \multicolumn{2}{c}{Aperture centres} & Observation  \\
  &    RA (J2000) & Dec. (J2000)                 &  RA (J2000) & Dec. (J2000) &  date \\
  &  (h m s)      & ($\degr$ $\arcmin$ $\arcsec$)        & (h m s)      & ($\degr$ $\arcmin$ $\arcsec$)  &        \\
\hline
G107.7$-$5.1:            &                  &                    & &   &            \\

 NW1   & 22 58 26.9  &   +55 16 02.3   & 22 58 27.5   & +55 16 02.3         &    2025 Sep 21    \\
 NW2   & 23 00 05.2  &   +55 28 57.9   & 23 00 05.7   & +55 28 57.9         &    2025 Aug 24    \\
 NW3   & 22 59 14.2  &   +55 18 48.1   & 22 59 12.3   & +55 18 48.1         &    2025 Aug 29    \\
 E1    & 23 08 33.9  &   +54 49 01.4   & 23 08 35.4   & +54 49 01.4         &    2025 Aug 27    \\
 S1    & 23 00 33.6  &   +53 39 04.3   & 23 00 32.6   & +53 39 04.3         &    2025 Sep 22     \\

\hline
G150.3+4.5:               &                  &                       & &   &           \\
   
 S1   & 04 26 02.5    &  +54 32 47.0  &04 26 03.5    &  +54 32 47.0          &  2025 Sep 21   \\
 S2   & 04 23 07.1    &  +54 31 12.1  & 04 23 04.9    &  +54 31 12.1         &  2025 Sep 22   \\
 S3   & 04 22 51.9   &  +54 32 15.9  & 04 22 51.6   &  +54 32 15.9         &  2025 Sep 23   \\
 S4   & 04 24 47.9   &  +54 39 03.2  & 04 24 45.1   &  +54 39 03.2          &  2025 Sep 23   \\
 S5   & 04 27 30.0   &  +54 40 30.1  & 04 27 30.2   &  +54 40 30.1          &  2025 Sep 24   \\
 S6   & 04 24 12.1   &  +54 29 56.4  &  04 24 09.3   &  +54 29 56.4          &  2025 Sep 25   \\  
 S7   & 04 23 24.5   &  +54 29 18.8  & 04 23 25.3   &  +54 29 18.8         &  2025 Sep 25   \\
 S8   & 04 27 10.0   &  +54 40 29.3   & 04 27 13.1   &  +54 40 29.3          &  2025 Sep 26   \\
\hline
\end{tabular}
\end{threeparttable}
\end{table*}
%%%%%%%%%%%%%%%%%%%%%%%%%%%%%%%%%%%%%%%%%%%%%%%%%%%%%%%%%%

\section{Analysis and Results}
\label{sec:analysis}
\subsection{H$\alpha$ images}
\label{sec:imaging}
\begin{figure*}
\includegraphics[angle=0, width=13cm]{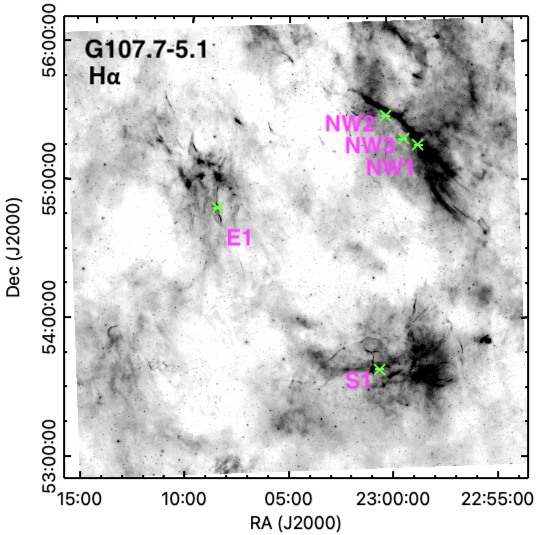}
\caption{H$\alpha$ image ($\sim$3.2 $\times$ 3.2 deg$^{2}$) of G107.7$-$5.1 from \citet{Fesen2024}, revealing well-defined filamentary emission distributed across the northwestern, eastern, and southern regions. The slit positions (NW1, NW2, NW3, E1, and S1) for the spectroscopy are marked with green lines, and the aperture centres of each slit are indicated by crosses.}
\label{fig:figure1}
\end{figure*}

\begin{figure*}
\includegraphics[angle=0, width=14cm]{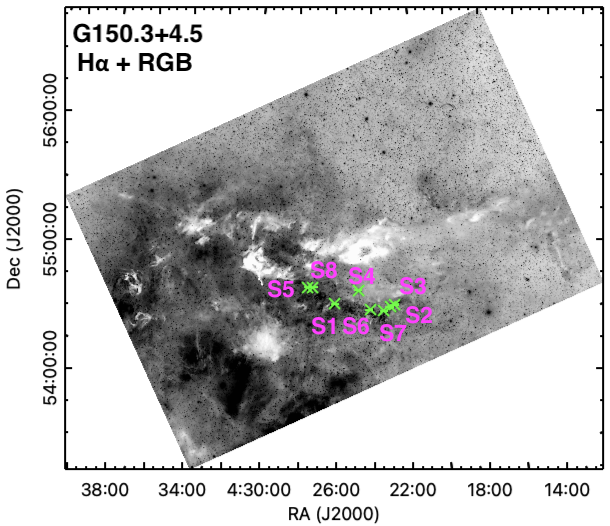}
\caption{H$\alpha$ plus RGB image ($\sim$3.5 $\times$ 2.4 deg$^{2}$) from \citet{Fesen2024}, showing prominent filaments along the southern region of G150.3+4.5. The slit positions (S1$-$S8) for the spectroscopy are marked with green lines, and the aperture centres of each slit are indicated by crosses.} 
\label{fig:figure2}
\end{figure*}

\citet{Fesen2024} presented high-quality images that provided valuable information for identifying G107.7$-$5.1 and G150.3+4.5 as SNRs in optical band. We present the H$\alpha$ image of G107.7$-$5.1 and the H$\alpha$+RGB composite image of G150.3+4.5 in Figs~\ref{fig:figure1} and \ref{fig:figure2}, respectively, taken from \citet{Fesen2024}. The images show that the emission associated with G107.7$-$5.1 is concentrated in its northwestern, eastern, and southern regions (see Fig.~\ref{fig:figure1}), whereas for G150.3+4.5, the emission is primarily confined to the southern part of the remnant (see Fig.~\ref{fig:figure2}).

%%%%%%%%%%%%%%%%%%%%%%%%%%%%%%%%%%%%%%%%%%%%%%%%%%%%%%%%%%
%%%% Figure 1: Image of G107.7$-$5.1
%%%%%%%%%%%%%%%%%%%%%%%%%%%%%%%%%%%%%%%%%%%%%%%%%%%%%%%%%%

%%%%%%%%%%%%%%%%%%%%%%%%%%%%%%%%%%%%%%%%%%%%%%%%%%%%%%%%%%
%%%% Figure 2: Image of G150.3+4.5

%%%%%%%%%%%%%%%%%%%%%%%%%%%%%%%%%%%%%%%%%%%%%%%%%%%%%%%%%%
\subsection{Optical long-slit spectra}
\label{sec:spectra}

To examine the nature and shock characteristics of the SNRs G107.7$-$5.1 and G150.3+4.5, long-slit spectra covering 4500$-$7000 {\AA} were obtained at multiple locations. We focused on regions exhibiting bright H$\alpha$ filamentary morphology, as reported by \citet{Fesen2024}. 

For G107.7$-$5.1, five positions located in the northwestern, eastern, and southern parts of the remnant were selected for spectroscopic observations. These positions are marked with crosses in Fig.~\ref{fig:figure1} and are labelled as NW1, NW2, and NW3 (northwestern positions), E1 (eastern position), and S1 (southern position). 

For G150.3+4.5, spectra were collected from eight positions within the southern filaments, indicated by crosses in Fig.~\ref{fig:figure2} and labelled as S1$-$S8. 

\begin{figure*}
\includegraphics[angle=0, width=8.2cm]{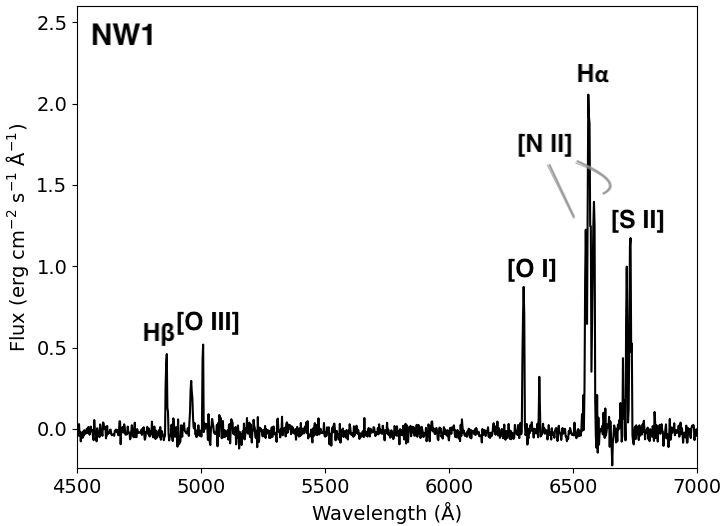}
\includegraphics[angle=0, width=8.2cm]{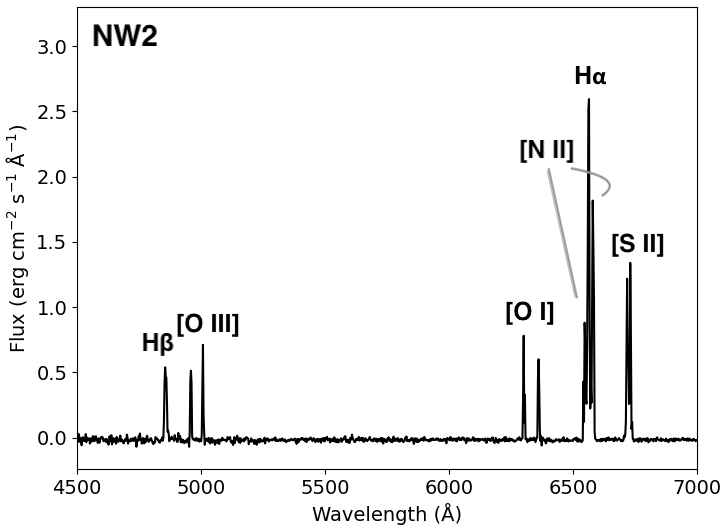}
\includegraphics[angle=0, width=8.2cm]{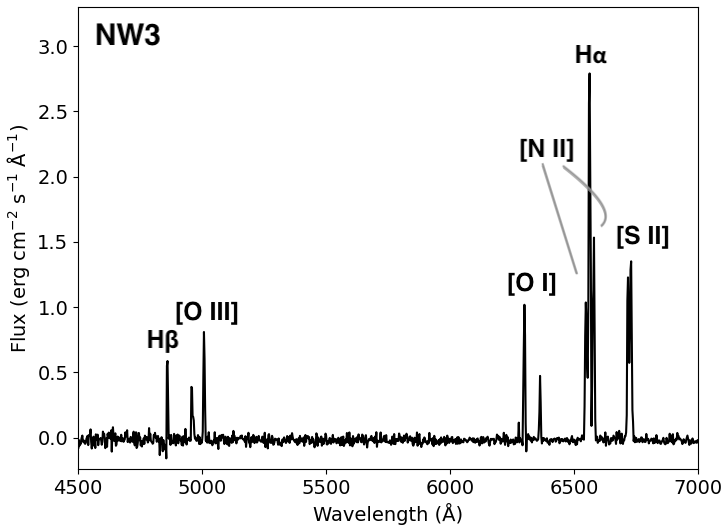}
\includegraphics[angle=0, width=8.2cm]{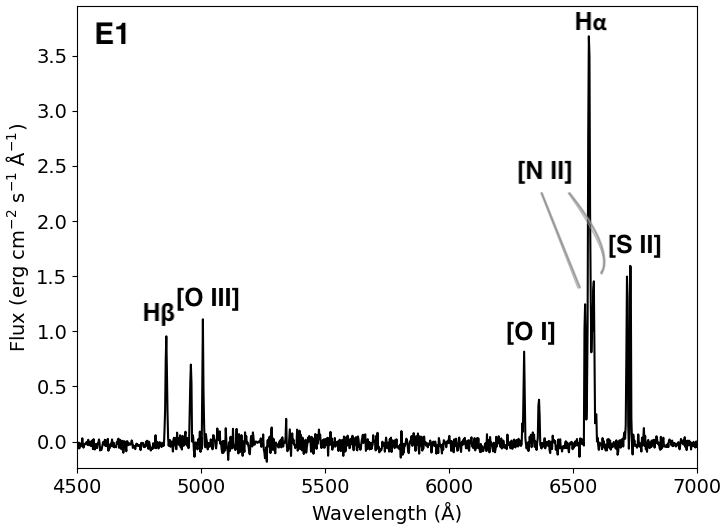}
\includegraphics[angle=0, width=8.2cm]{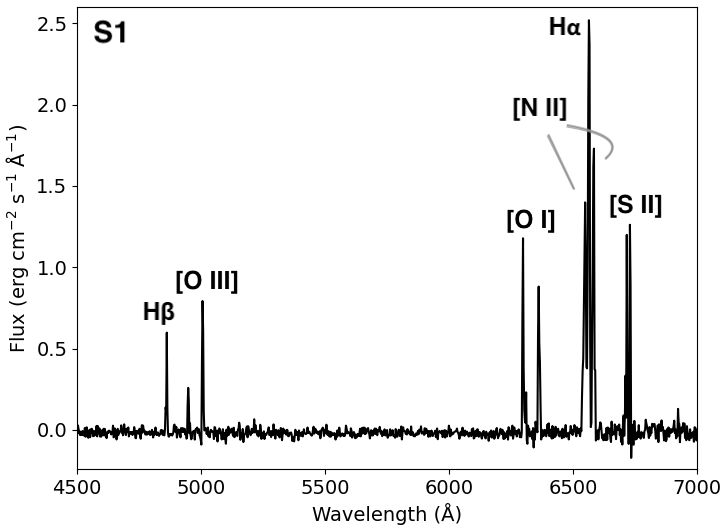}
\caption{RTT150 long-slit spectra obtained at three positions in the northwestern region (NW1, NW2, and NW3), one position in the eastern region (E1), and one position in the southern region (S1) of G107.7$-$5.1, covering the wavelength range 4500$-$7000 {\AA}. Fluxes are in $10^{-16}$ erg cm$^{-2}$ s$^{-1}$ \AA$^{-1}$.}
\label{fig:figure3}
\end{figure*}

\begin{figure*}
\includegraphics[angle=0, width=7.9cm]{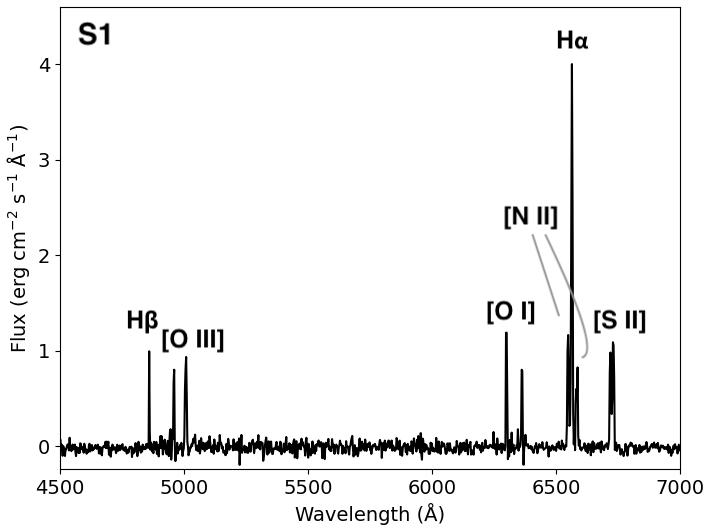}
\includegraphics[angle=0, width=7.9cm]{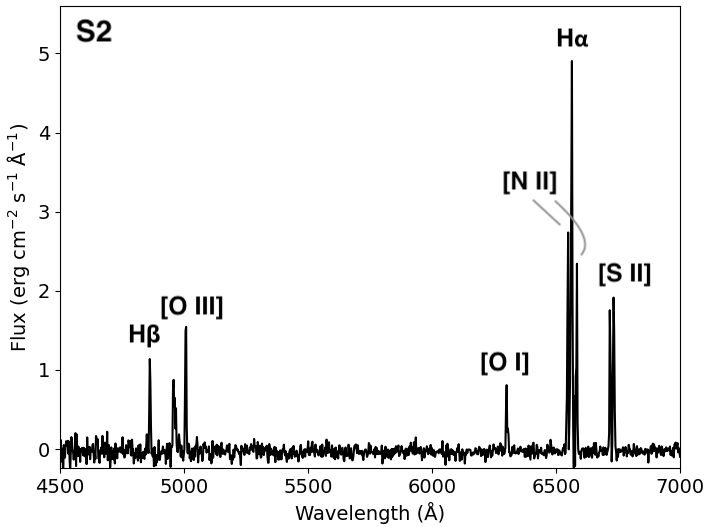}
\includegraphics[angle=0, width=7.9cm]{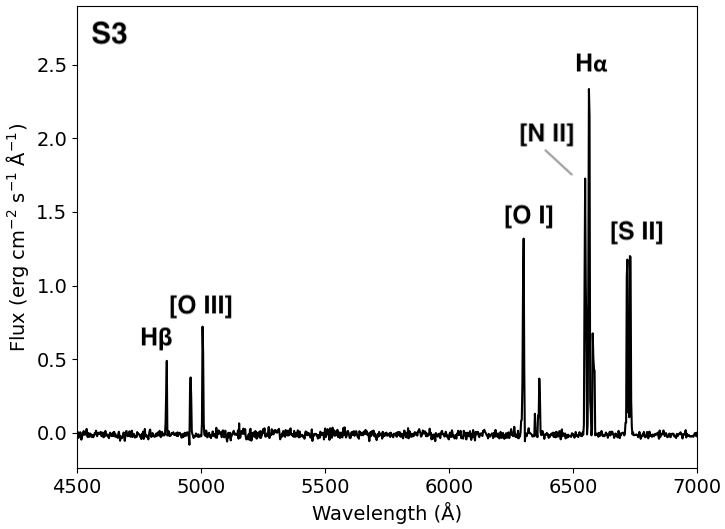}
\includegraphics[angle=0, width=7.9cm]{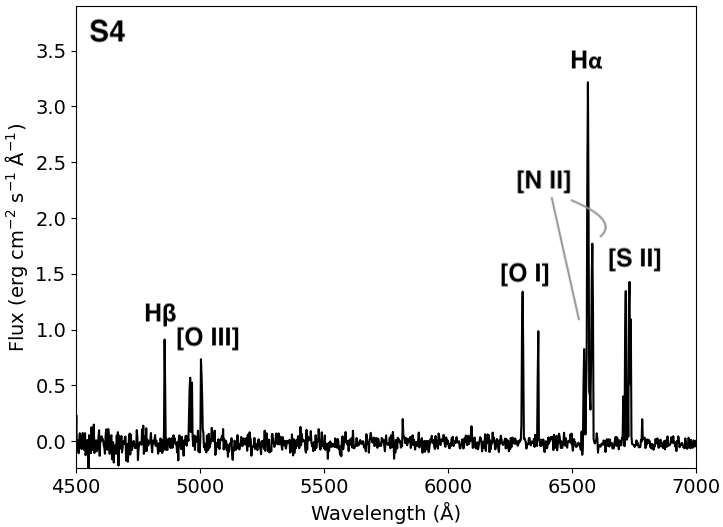}
\includegraphics[angle=0, width=7.9cm]{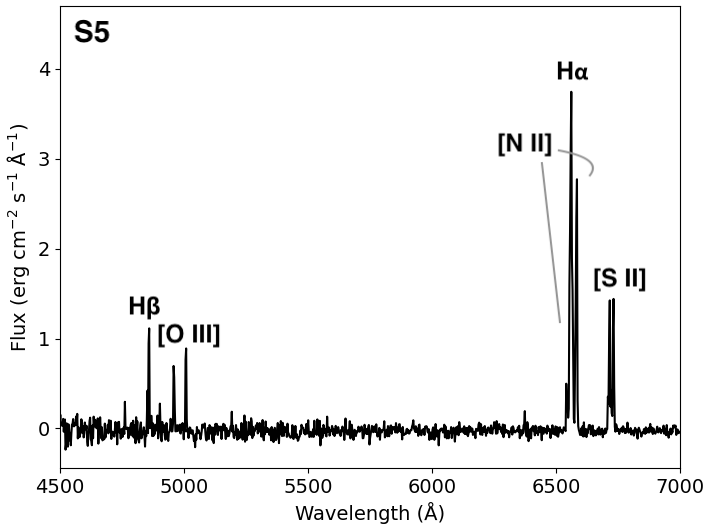}
\includegraphics[angle=0, width=7.9cm]{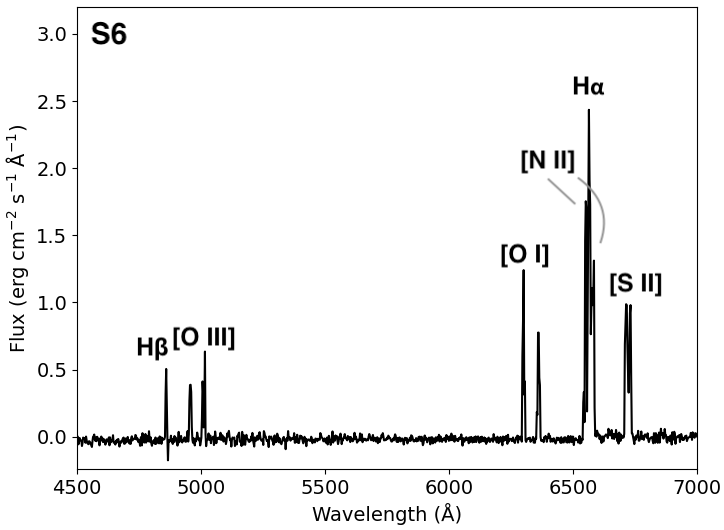}
\includegraphics[angle=0, width=7.9cm]{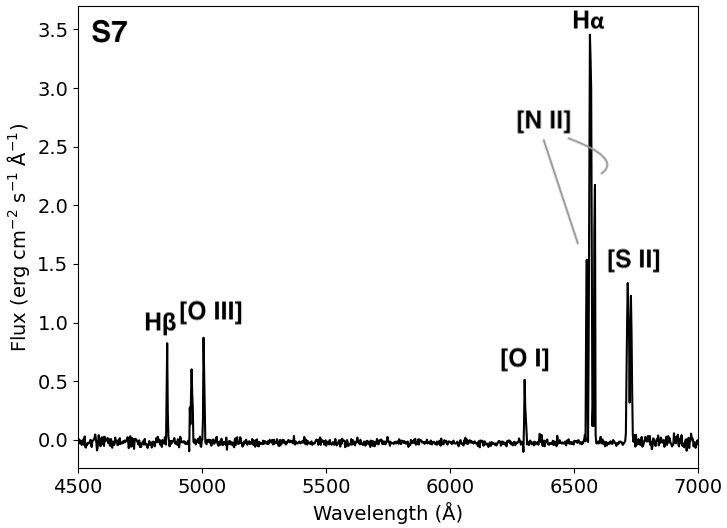}
\includegraphics[angle=0, width=7.9cm]{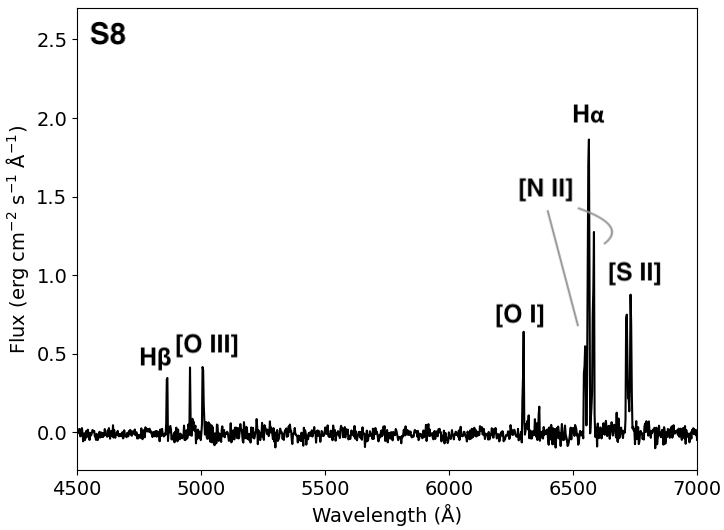}
\caption{RTT150 long-slit spectra obtained at positions S1$-$S8 in the southern region of G150.3+4.5, covering the wavelength range 4500$-$7000 {\AA}. Fluxes are in $10^{-16}$ erg cm$^{-2}$ s$^{-1}$ \AA$^{-1}$.}
\label{fig:figure4}
\end{figure*}

Spectra of both SNRs are shown in Figs~\ref{fig:figure3} and \ref{fig:figure4}, respectively. We detected emission from H$\alpha$ $\lambda$6563, H$\beta$ $\lambda$4861, [\ion{O}{iii}], [\ion{O}{i}], [\ion{N}{ii}], and [\ion{S}{ii}]. The fluxes of strong emission lines and uncertainties are calculated by the Gaussian-profile fitting with a \textsc{python} package following \citet{Bakis2025a}. The 1-$\sigma$ uncertainties of the emission-line fluxes were obtained  from the curve-fitting procedure applied to each spectral line. We derive the key diagnostic line ratios, [\ion{S}{ii}]/H$\alpha$, [\ion{N}{ii}]/H$\alpha$, [\ion{S}{ii}] $\lambda$6716/$\lambda$6731, and [\ion{O}{iii}]/H$\beta$, and estimate the associated physical parameters. The uncertainties of the line ratios were derived using standard 
error propagation based on the differentiation of the corresponding 
flux ratios. The parameter uncertainties correspond to 1-$\sigma$ error estimates derived from the covariance matrix and scaled by the reduced $\chi^{2}$. The uncertainties of parameter ratios were calculated using standard error propagation, taking into account the covariance between the fitted parameters. The results of our spectroscopic analysis are presented in Tables~\ref{tab:Table2} and \ref{tab:Table3}, where the observed emission-line fluxes, $F$, are given on a scale where $F$(H$\alpha$) = 100.

%%%%%%%%%%%%%%%%%%%%%%%%%%%%%%%%%%%%%%%%%%%%%%%%%%%%%%%%%%
%%%% Figure 3: Long-slit spectra of G107.7$-$5.1
%%%%%%%%%%%%%%%%%%%%%%%%%%%%%%%%%%%%%%%%%%%%%%%%%%%%%%%%%%

%%%% Figure 4: Long-slit spectra of G150.3+4.5
%%%%%%%%%%%%%%%%%%%%%%%%%%%%%%%%%%%%%%%%%%%%%%%%%%%%%%%%%%

%%%% Table 2: Line fluxes for G107.7$-$5.1
\begin{table*}
\centering
\caption{Integrated line fluxes for G107.7$-$5.1, shown relative to $F$(H$\alpha$) = 100, with 1$\sigma$ uncertainties. Diagnostic line ratios and the derived physical parameters are additionally provided. Details of the uncertainty estimates are described in Section~\ref{sec:spectra}.}
\label{tab:Table2}
 \begin{threeparttable}
 \begin{tabular}{@{}p{3.2cm}p{1.8cm}p{1.8cm}p{1.8cm}p{1.8cm}p{1.8cm}@{}}
 \hline
 Slit ID:  &		NW1 	&	NW2 &	NW3 &	E1 &	S1   \\[0.5 ex]
 Region:  &		Northwestern	&	Northwestern  &	Northwestern  &	Eastern &	Southern   \\[0.5 ex]
 (Observation date):                             &(21Sep2025)         &(24Aug2025) &	(29Aug2025) &	(27Aug2025) &	(22Sep2025)   \\[0.5 ex]
  
\hline
 Lines &			&	  &	  &	 &	   \\[0.5 ex]
H$\beta$ $\lambda$4861	        &	$14\pm2$	&	$31\pm3$	& $19\pm6$  & $21\pm1$ & $10\pm2$	  \\
												
[\ion{O}{iii}] $\lambda$4959 	&	$13\pm2$	&	$21\pm4$	&  $11\pm4$      & $20\pm4$ &	$9\pm3$	   \\
												
[\ion{O}{iii}] $\lambda$5007 	&	$9\pm3$	   &	$16\pm2$    &  $22\pm2$      &  $17\pm3$ &	$25\pm4$	  \\
												
[\ion{O}{i}] $\lambda$6300 		&	$28\pm4$	&	$16\pm3$	& $34\pm5$        & $15\pm3$ &	$30\pm6$	 \\
												
[\ion{O}{i}] $\lambda$6363 		&	$6\pm1$	    &	$22\pm5$	&  $11\pm2$       & $9\pm2$ &	$38\pm7$	 \\
												
[\ion{N}{ii}] $\lambda$6548 	&	$23\pm8$	&	$33\pm13$	  &   $36\pm8$    & $25\pm5$ &	 $73\pm14$	  \\
												
H$\alpha$ $\lambda$6563 	    &	$100\pm15$	   &	$100\pm12$  &   $100\pm8$    & $100\pm7$ &	$100\pm11$	 \\
												
[\ion{N}{ii}] $\lambda$6584     &	$47\pm10$	    &	$72\pm12$       &   $50\pm8$    & $53\pm8$ &	$67\pm11$	 \\
												
[\ion{S}{ii}] $\lambda$6716 	&	$23\pm5$	    &	$44\pm7$        & $36\pm4$  & $30\pm4$ &	$25\pm8$	   \\
												
[\ion{S}{ii}] $\lambda$6731 	&	$38\pm9$	    &	$31\pm5$ 	    & $50\pm5$  & $26\pm4$ &	$34\pm11$		\\

$F$(H$\alpha$)$^{\rm a}$      &	$31.1\pm4.7$	&	$21.8\pm2.7$	& $26.7\pm2.2$  & $38.0\pm2.6$ & $22.8\pm2.5$	 \\
\hline

 Line Ratios &			&	  &	  &	 &	   \\[0.5 ex]

[\ion{S}{ii}]$^{\rm b}$/H$\alpha$ 	&	  $0.60\pm0.01$  &	   $0.75\pm0.06$  & $0.86\pm0.01$ & $0.56\pm0.05$ &	$0.59\pm0.10$	 \\

[\ion{N}{ii}]$^{\rm c}$/H$\alpha$ 	&	   $0.70\pm0.08$ &	    $1.05\pm0.13$  &  $0.86\pm0.08$ & $0.78\pm0.08$ & $1.40\pm0.09$		 \\

[\ion{O}{iii}]$\lambda$5007/H$\beta$ 	& $0.64\pm0.09$	  &	 $0.52\pm0.01$   & $1.16\pm0.27$  & $0.81 \pm0.10$ & $2.50\pm0.01$		 \\

[\ion{O}{iii}]$^{\rm d}$/H$\beta$ 	& $1.64\pm 0.08$  &	 $1.19\pm 0.06 $  & $1.76\pm 0.23 $ & $1.73\pm 0.23 $& 	$3.33\pm 0.12	$ \\

[\ion{S}{ii}]$\lambda$6716/$\lambda$6731 &	  $0.60\pm0.01$  &$1.42\pm0.01$ & $0.73\pm0.01$ & $1.14\pm0.02$ &	$0.75\pm0.01$ \\	
\hline 																				 Parameters &			&	  &	  &	 &	   \\[0.5 ex]	
$n_{\rm e}$(cm$^{-3}$)	          &	 $4650\pm650$  & $30\pm1$   & $2180\pm300$ & $400\pm40$ & $1990\pm190$	 \\

$n_{\rm 0}$ (cm$^{-3}$) & $100 \pm 10$ & $0.7 \pm 0.02$ & $50 \pm 10$ & $9 \pm 1$ & $40 \pm 4$    \\
$c$(H$\beta$)$^{\rm e}$                               &   $1.14 \pm0.19$ &	   $0.10 \pm 0.13$  &   $0.74\pm0.41$    & $0.61\pm0.06$ & $1.58 \pm0.26$    \\
$E(B-V)$                          &   $0.88\pm0.15$  &	   $0.07 \pm0.10$    &  $0.57\pm0.32$     & $0.47\pm0.05$ &	$1.22\pm0.20$      \\
$A_{\rm v}$                       &   $2.72 \pm0.45$ &	   $0.23 \pm0.30$    &  $1.76\pm0.99$     & $1.45 \pm0.15$&	$3.77\pm0.63$       \\
$N_{\rm H}$ ($10^{21}$ cm$^{-2}$) &   $4.73\pm0.78$  &	   $0.40\pm0.53$    & $3.07\pm1.72$      & $2.52\pm0.26$ &	$6.57 \pm1.09$   \\
\hline  

\end{tabular}
\begin{tablenotes}
\item {{\it Notes.} $^{\rm a}$ Fluxes in units of $10^{-16}$ erg cm$^{-2}$ s$^{-1}$ \AA$^{-1}$. \\
$^{\rm b}$ Sum of the $\lambda$6716 and $\lambda$6731 line fluxes. \\
$^{\rm c}$ Sum of the $\lambda$6548 and $\lambda$6584 line fluxes. \\
$^{\rm d}$ Sum of the $\lambda$4959 and $\lambda$5007 line fluxes. \\
$^{\rm e}$ The logarithmic extinction is calculated by $c$=1/0.331 $\times$ log[(H$\alpha$/H$\beta$)/3].
} 
\end{tablenotes}
\end{threeparttable}
\end{table*}
%%%%%%%%%%%%%%%%%%%%%%%%%%%%%%%%%%%%%%%%%%%%%%%%%%%%%%%%%%

%%%%%%%%%%%%%%%%%%%%%%%%%%%%%%%%%%%%%%%%%%%%%%%%%%%%%%%%%%
%%%% Table 3: Line fluxes for G150.3+4.5
\begin{table*}
\centering
\caption{Same as Table~\ref{tab:Table2}, but for G150.3+4.5.}
\label{tab:Table3}
 \begin{threeparttable}
 \begin{tabular}{@{}p{3.2cm}p{1.8cm}p{1.8cm}p{1.8cm}p{1.8cm}@{}}
 \hline
      Slit ID:           &		S1	      &	S2 	          & S3            & S4 	 \\[0.5 ex]
	Region:		         &	 Southern	    &	Southern 	& Southern  	& Southern  \\[0.5 ex]			
    (Observation date)	 &   (22Sep2025)	&	(21Sep2025) & (23Sep2025)  	& (23Sep2025) \\[0.5 ex]			
               \hline	
Lines &			&	  &	  &	 	   \\[0.5 ex]
H$\beta$ $\lambda$4861	        &	$12\pm3$	&	$20\pm2$	&	$13\pm5$	& $17\pm3$	    \\
												
[\ion{O}{iii}] $\lambda$4959 	&	$20\pm8$	&	$27\pm9$	&	$21\pm5$	&	$18\pm11$    \\
												
[\ion{O}{iii}] $\lambda$5007 	&	$35\pm4$	&	$31\pm4$	&	$23\pm2$	&	$23\pm6$  \\
												
[\ion{O}{i}] $\lambda$6300 		&	$34\pm6$	&	$16\pm5$	&	$56\pm7$	& $37\pm3$	  \\
												
[\ion{O}{i}] $\lambda$6363 		&	$25\pm8$	&	$...$	&	$16\pm5$	&	$17\pm4$   \\
												
[\ion{N}{ii}] $\lambda$6548 	&	$37\pm7$	&	$43\pm16$	&	$76\pm16$	&	$27\pm9$  \\
												
H$\alpha$ $\lambda$6563 	    &	$100\pm6$	&	$100\pm19$   &	$100\pm15$	& $100\pm10$  \\
												
[\ion{N}{ii}] $\lambda$6584     &	$30\pm8$	&	$40\pm18$	&	$38\pm18$	&	$64\pm10$  \\
												
[\ion{S}{ii}] $\lambda$6716 	&	$26\pm4$	&	$32\pm6$	&	$43\pm8$	 &  $32\pm14$  \\
												
[\ion{S}{ii}] $\lambda$6731 	&	$38\pm6$	&	$40\pm7$	&	$47\pm8$	&	$35\pm15$ 		\\

$F$(H$\alpha$)$^{\rm a}$      &	$26.6\pm1.7$&	$37.3\pm7.1$ &	$15.2\pm2.2$	& $25.5\pm2.5$ 	 \\
\hline 

 Line ratios &			&	  &	  &	 	   \\[0.5 ex]
[\ion{S}{ii}]/H$\alpha$ 	   &  $0.63\pm0.04$ &	$0.72\pm0.02$ &	$0.90\pm0.02$	& $0.67\pm0.21$  \\

[\ion{N}{ii}]/H$\alpha$ 	   &	$0.66\pm0.11$&	$0.83\pm0.18$ &	$1.14\pm0.18$	& $0.91\pm0.11$  \\

[\ion{O}{iii}]$\lambda$5007/H$\beta$ 	&  $2.92\pm0.39$ &	$1.21\pm0.11$   & $1.75\pm0.55$  & $1.36\pm0.12$ 	 \\

[\ion{O}{iii}]/H$\beta$ 	  & 	 $4.63\pm 0.21	$ &		$2.86\pm 0.32 $  &   $3.32\pm  0.75$ & 	   $2.38\pm  0.58	$ \\

[\ion{S}{ii}]$\lambda$6716/$\lambda$6731 &	   $0.68\pm0.01$  &	   $0.80\pm0.01$ &	$0.91\pm0.02$ & $0.89\pm0.01$  \\	
\hline 
Parameters     &			&	  &	  &	 	   \\[0.5 ex]

$n_{\rm e}$ (cm$^{-3}$) & $2800 \pm 200$ & $1600 \pm 200$ & $1100 \pm 200$ & $1100 \pm 200$  \\

$n_{\rm 0}$ (cm$^{-3}$) & $60 \pm 4$ & $40 \pm 4$ & $20 \pm 4$ & $20 \pm 4$   \\
$c$(H$\beta$)                               &   $1.34\pm 0.33$  &	   $0.67\pm0.13$    &	$1.24\pm0.51$   & $0.88\pm0.23$     \\
$E(B-V)$                          &   $1.03\pm0.25$   &	   $0.52\pm0.10$    &	$0.95\pm0.39$   & $0.68\pm0.18$    \\
$A_{\rm v}$                       &   $3.2\pm0.78$    &	   $1.60 \pm0.31$   &	$2.95\pm1.21$   & $2.11 \pm0.55$   \\
$N_{\rm H}$ ($10^{21}$ cm$^{-2}$) &   $5.57 \pm1.36$  &	   $2.79\pm0.55$    &	$5.14\pm2.10$   & $3.67\pm0.96$    \\
\hline  
  &		S5	&	S6 	& S7  & S8  	 \\[0.5 ex]
			   &		Southern	&	Southern 	& Southern  	& Southern \\[0.5 ex]								  &		                     (24Sep2025)	&	(25Sep2025) 	& (25Sep2025) 	& (26Sep2025)  \\[0.5 ex]				
               \hline	
Lines &			&	  &	  &	 	   \\[0.5 ex]
H$\beta$ $\lambda$4861	        &	$13\pm7$	&   $17\pm7$		&	$11\pm1$	& $14\pm4$	  \\
												
[\ion{O}{iii}] $\lambda$4959 	&	$10\pm2$	&	$18\pm3$	&	$18\pm4$	&	$10\pm3$   \\
												
[\ion{O}{iii}] $\lambda$5007 	&	$13\pm3$	&	$23\pm11$	&	$14\pm2$	&	$21\pm4$  \\
												
[\ion{O}{i}] $\lambda$6300 		&	$...$	&	$38\pm9$	&	$12\pm3$	&	$24\pm7$ \\
												
[\ion{O}{i}] $\lambda$6363 		&	$...$	&	$32\pm6$	&	$...$	&	$6\pm6$ \\
												
[\ion{N}{ii}] $\lambda$6548 	&	$9\pm9$	&	$59\pm14$	&	$21\pm5$	& $38\pm16$	  \\
												
H$\alpha$ $\lambda$6563 	    &	$100\pm12$	&	$100\pm21$   &	$100\pm8$	&	$100\pm14$ \\
												
[\ion{N}{ii}] $\lambda$6584     &	$50\pm9$	&	$70\pm23$	&	$28\pm12$	&$64\pm14$	 \\
												
[\ion{S}{ii}] $\lambda$6716 	&	$21\pm6$	&	$54\pm6$	&	$37\pm5$	 &  $46\pm10$  \\
												
[\ion{S}{ii}] $\lambda$6731 	&	$22\pm6$	&	$38\pm5$	&	$26\pm4$	&	$44\pm9$		\\

$F$(H$\alpha$)$^{\rm a}$      &	$41.0\pm4.8$	&	$23.9\pm4.9$	&	$38.6\pm3.0$	& $16.0\pm2.2$	 \\
\hline 
Line ratios &			&	  &	  &	 	   \\[0.5 ex]
[\ion{S}{ii}]/H$\alpha$ 	&	  $0.43\pm0.07$  &	   $0.92\pm0.06$ &	$0.63\pm0.05$	& $0.91\pm0.07$ \\

[\ion{N}{ii}]/H$\alpha$ 	&	   $0.59\pm0.11$  &	   $1.29\pm0.11$ &	$0.49\pm0.13$	&  $1.02\pm0.15$\\

[\ion{O}{iii}]$\lambda$5007/H$\beta$ 	& $1.01\pm0.33$  &	$1.31\pm0.14$   &  $1.28\pm0.04$ &   $1.48\pm0.13$		 \\

[\ion{O}{iii}]/H$\beta$ 	  & 	$ 1.76\pm 0.60$ &	$2.38\pm 0.58$   & 	$ 2.94 \pm  0.23$    & 	 $2.18 \pm  0.90$	 \\

[\ion{S}{ii}]$\lambda$6716/$\lambda$6731 &	   $0.99\pm0.01$  &	   $1.41\pm0.01$ &	$1.40\pm0.03$ & $1.05\pm0.01$ \\	
\hline 																			Parameters     &			&	  &	  &	 	   \\[0.5 ex]		
$n_{\rm e}$ (cm$^{-3}$) & $740 \pm 70$ & $40 \pm 1$ & $50 \pm 1$ & $580 \pm 20$  \\

$n_{\rm 0}$ (cm$^{-3}$) & $20 \pm 2$ & $0.9 \pm 0.02$ & $1.1 \pm 0.02$ & $10 \pm 1$  \\
$c$(H$\beta$)                               &   $1.24\pm0.71$       &	   $0.88\pm0.54$    &	$1.45\pm0.12$   &  $1.14\pm0.38$ \\
$E(B-V)$                          &   $0.95\pm0.54$       &	   $0.68\pm0.42$    &	$1.12 \pm0.09$  &  $0.88\pm0.29$\\
$A_{\rm v}$                       &   $2.95\pm1.68$       &	   $2.11 \pm1.29$   &	$3.47 \pm0.29$  &  $2.72 \pm0.90$ \\
$N_{\rm H}$ ($10^{21}$ cm$^{-2}$) &  $ 5.13\pm2.94 $      &	   $3.67\pm2.25$    &	$6.05\pm0.50$   & $ 4.73\pm1.56$ \\
\hline  

\end{tabular}
\begin{tablenotes}
\item {{\it Note.} $^{\rm a}$ Fluxes in units of $10^{-16}$ erg cm$^{-2}$ s$^{-1}$ \AA$^{-1}$.} 
\end{tablenotes}
\end{threeparttable}
\end{table*}
%%%%%%%%%%%%%%%%%%%%%%%%%%%%%%%%%%%%%%%%%%%%%%%%%%%%%%%%%%

%%%%%%%%%%%%%%%%%%%%%%%%%%%%%%%%%%%%%%%%%%%%%%%%%%%%%%%%%%
\clearpage\clearpage
\section{Discussion}
\label{sec:discuss}
Spectroscopic observations conducted with the TFOSC instrument on RTT150 telescope enabled us to investigate the Balmer and forbidden emission lines in the spectra of the SNRs G107.7$-$5.1 and G150.3+4.5.  We calculated the emission-line ratios ([\ion{S}{ii}]/H$\alpha$, [\ion{N}{ii}]/H$\alpha$, [\ion{S}{ii}] $\lambda$6716/$\lambda$6731, [\ion{O}{iii}]/H$\beta$, and H$\alpha$/H$\beta$; see Tables~\ref{tab:Table2} and \ref{tab:Table3}) and derived the relevant physical parameters. The resulting values for both SNRs are discussed below.

%%%%%%%%%%%%%%%%%%%%%%%%%%%%%%%%%%%%%%%%%%%%%%%%%%%%%%%%%%

\subsection{Optical spectral properties of G107.7$-$5.1}
We analysed the [\ion{S}{ii}]/H$\alpha$ and [\ion{N}{ii}]/H$\alpha$ line-ratio diagnostics commonly used for SNR identification to confirm the SNR nature of G107.7$-$5.1. The measured [\ion{S}{ii}]/H$\alpha$ ratios range from 0.56 to 0.86 (listed in Table~\ref{tab:Table2}), exceeding the canonical threshold of $\geq 0.4$ (e.g. \citealt{Fesen1985, Leonidaki2013, Long2017}) that distinguishes shock-excited gas from photoionized regions, indicating that the emission arises from shock-heated material. The [\ion{N}{ii}]/H$\alpha$ ratios at all observed positions (0.7$-$1.4) further support the presence of shock-excited gas (e.g. \citealt{Fesen1985, FrewParker2010}). Additionally, the spectra of the optical filaments in all observed regions display [\ion{O}{i}] $\lambda$6300 and $\lambda$6363 emissions, providing further confirmation of their shock-excited nature \citep{Fesen1985, Kopsacheili2020}.

\begin{figure*}
\includegraphics[angle=0, width=7.9cm]{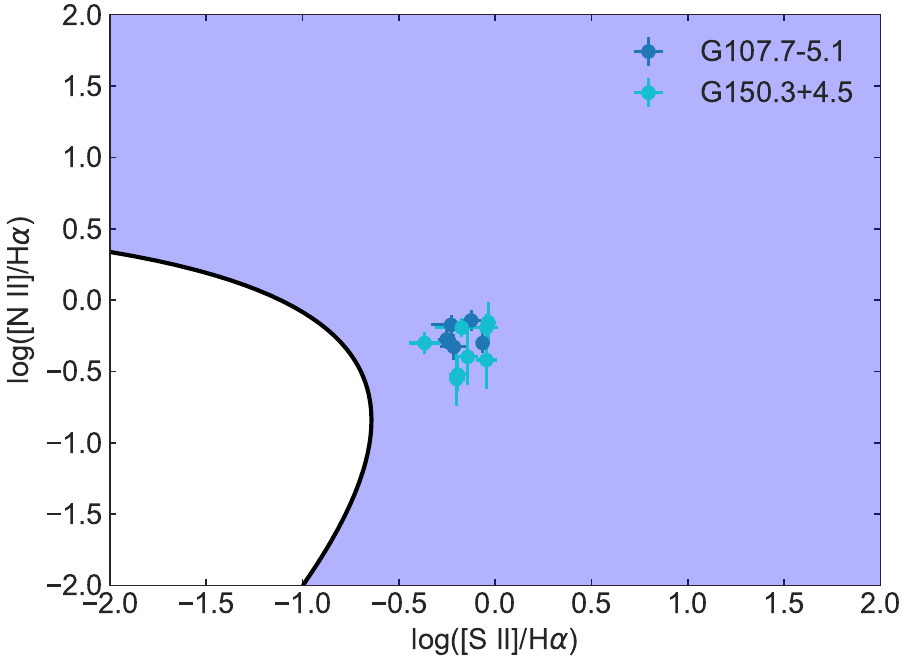}
\includegraphics[angle=0, width=7.9cm]{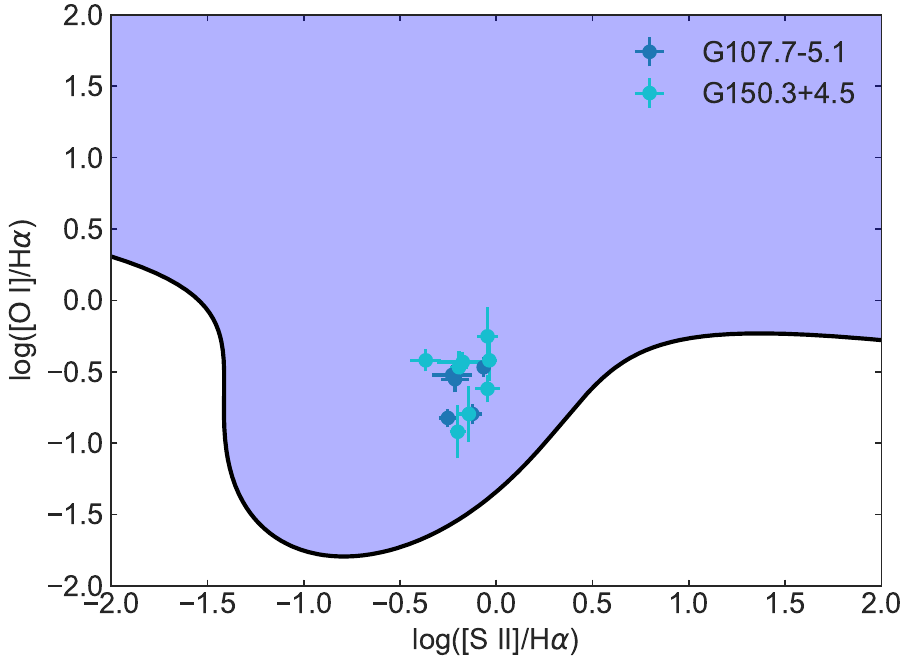}
\includegraphics[angle=0, width=7.9cm]{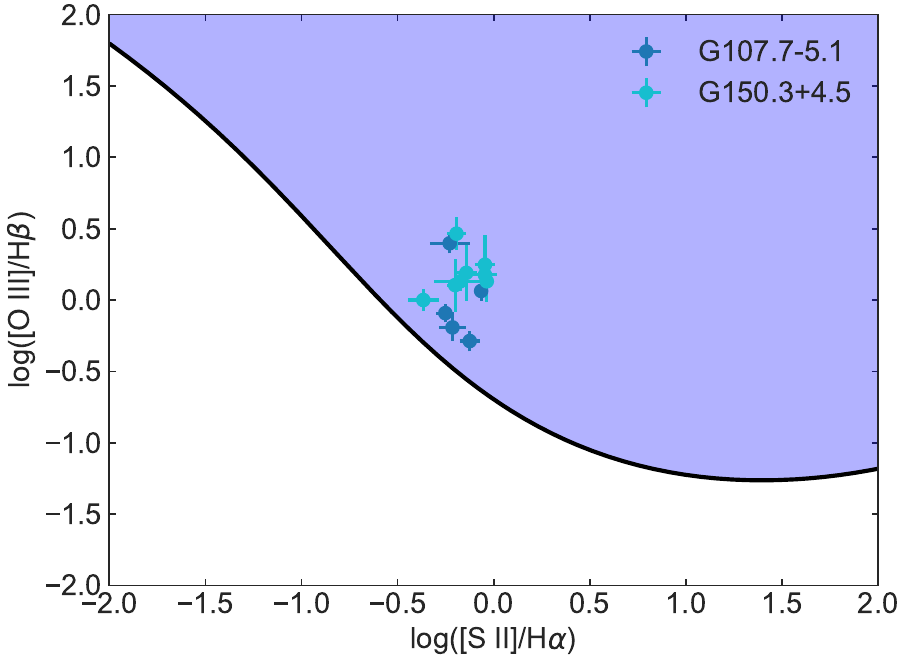}
\includegraphics[angle=0, width=7.9cm]{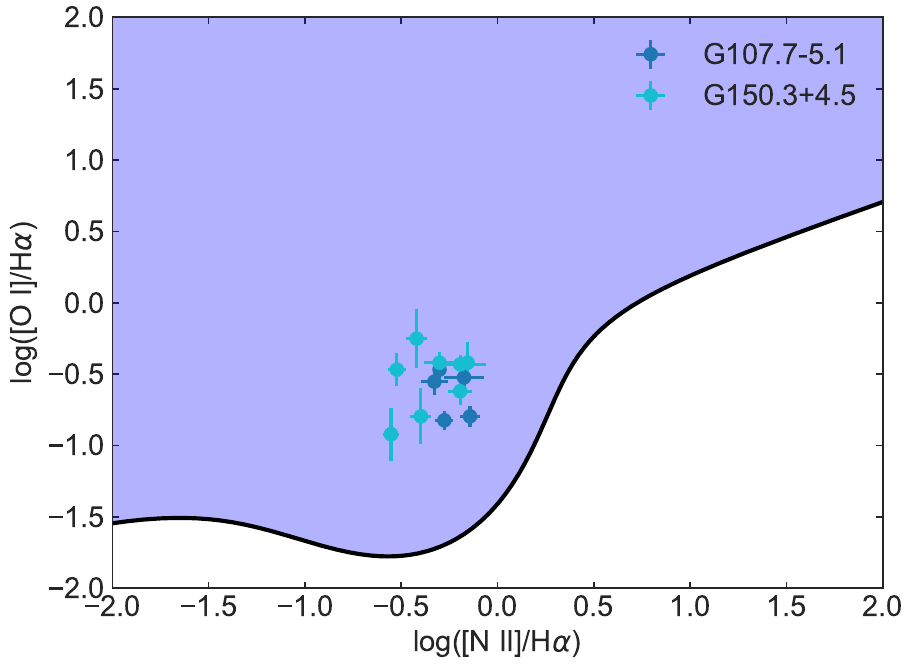}
\includegraphics[angle=0, width=7.9cm]{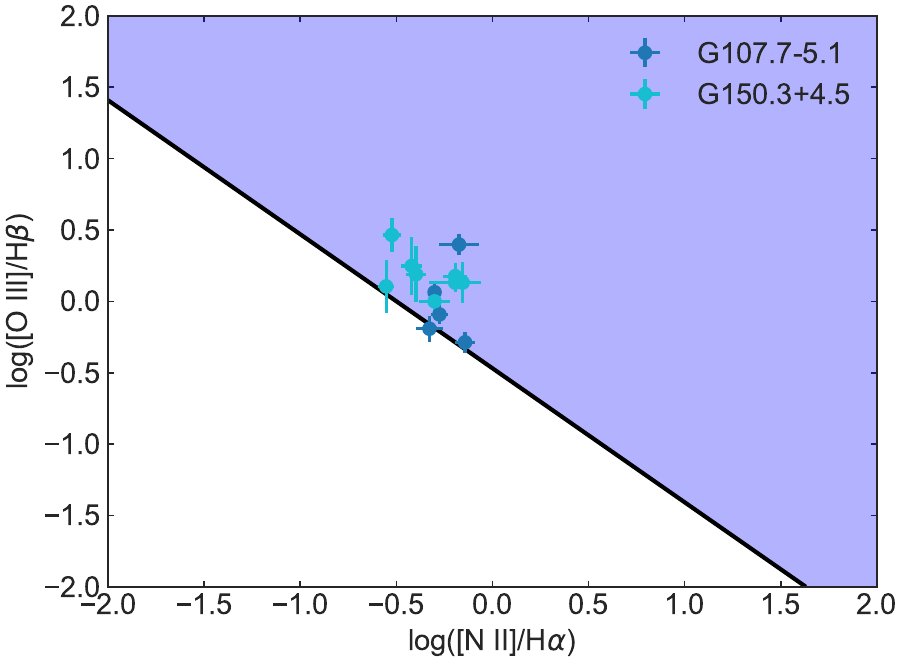}
\includegraphics[angle=0, width=7.9cm]{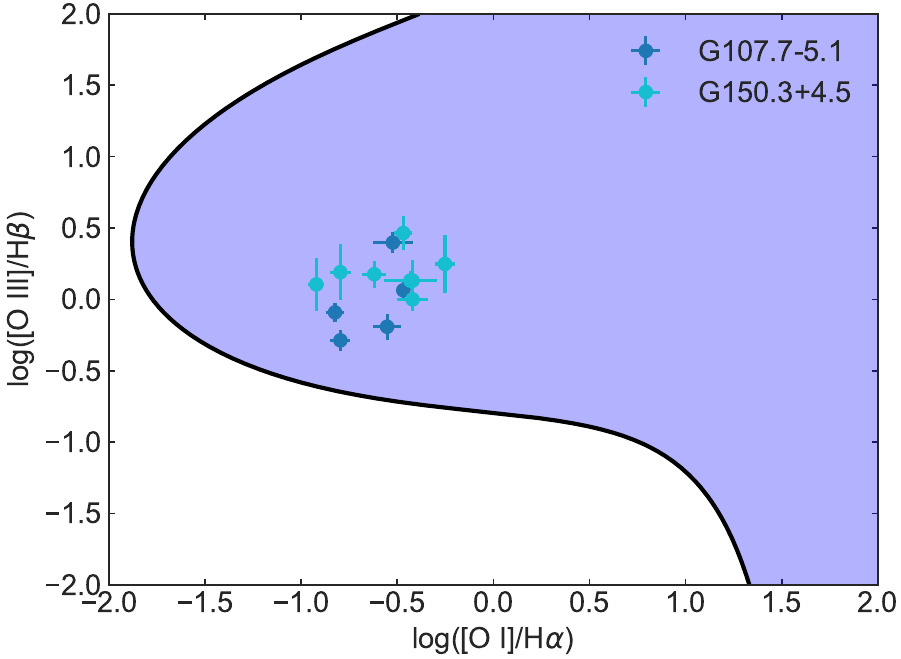}
\caption{2D diagnostic diagrams adapted from \citet{Kopsacheili2020}. All diagnostic data were obtained directly from the link provided by \citet{Kopsacheili2024} and are overlaid with our spectroscopic measurements of G107.7$-$5.1 (navy blue) and G150.3+4.5 (green). The blue shaded region denotes the parameter space characteristic of shock-excited gas.}
\label{fig:figure5}
\end{figure*}

In Fig.~\ref{fig:figure5}, we present diagnostic diagrams adapted from \citet{Kopsacheili2020}\footnote{The diagnostic data were obtained directly from the link made available by \citet{Kopsacheili2024}.}  that discriminate between shock-heated and photo-ionized regions based on different combinations of emission-line ratios: (1) [\ion{S}{ii}]/H$\alpha$ versus [\ion{N}{ii}] $\lambda$6584/H$\alpha$ (top left), (2) [\ion{S}{ii}]/H$\alpha$ versus [\ion{O}{i}] $\lambda$6300/H$\alpha$ (top right), (3)  [\ion{S}{ii}]/H$\alpha$ versus [\ion{O}{iii}] $\lambda$5007/H$\beta$ (middle left), (4) [\ion{N}{ii}] $\lambda$6584/H$\alpha$ versus [\ion{O}{i}] $\lambda$6300/H$\alpha$ (middle right), (5) [\ion{N}{ii}] $\lambda$6584/H$\alpha$ versus [\ion{O}{iii}] $\lambda$5007/H$\beta$ (bottom left), and (6) [\ion{O}{i}] $\lambda$6300/H$\alpha$ versus [\ion{O}{iii}] $\lambda$5007/H$\beta$ (bottom right). All ratios are presented on a logarithmic scale. For G107.7$-$5.1, the measured ratios obtained from all five positions fulfill the required criteria and occupy the parameter space characteristic of shock-heated regions.

%%%% Figure 5: diagnostic diagrams

The forbidden sulphur lines at $\lambda$6716 and $\lambda$6731 serve as key indicators of electron density \citep{OsterbrockFerland2006}. Since the [\ion{S}{ii}] $\lambda$6716 and $\lambda$6731 lines were detected in all spectra, the electron density ($n_{\rm e}$) could be derived. Using the \textsc{python}-based code \texttt{PyNeb}\footnote{\url{http://research.iac.es/proyecto/PyNeb/}}
 \citep{Luridiana2015} and adopting an electron temperature of $T_{\rm e}=10000$~K, we estimated $n_{\rm e}$ to be in the range of approximately 30–4650 cm$^{-3}$, based on the [\ion{S}{ii}] $\lambda$6716/$\lambda$6731 line ratio. The derived densities are generally consistent with those reported for Galactic SNRs, with the exception of region NW1. In this region, the [\ion{S}{ii}] $\lambda$6716/$\lambda$6731 ratio of 0.6 implies a comparatively high electron density of $\sim$4650 cm$^{-3}$. Conversely, the ratio measured at the NW2 position is close to the theoretical maximum value of $\sim$1.43 \citep{OsterbrockFerland2006}, consistent with a low electron density. We found significant variations in electron density among the northwestern, eastern, and southern regions of the remnant (see Table~\ref{tab:Table2}). These variations may reflect differing local ambient conditions, potentially including interactions with molecular material. However, no molecular gas has been detected in this region \citep{Araya2024}. Our results suggest that additional observations are necessary to determine whether the SNR is interacting with dense ISM material. Another possible explanation for the significant density variations may be the presence of dense circumstellar medium (CSM), similar to that observed in several SNRs in the Large Magellanic Cloud (LMC). In the case of N103B, the electron densities of the nebular knots 
 were found to range from $\sim$500 to 5000 cm$^{-3}$ based on the [\ion{S}{ii}] $\lambda$6716/$\lambda$6731 line ratio, supporting a circumstellar origin \citep{Li2017}. Subsequently, \citet{Li2021} reported numerous bright forbidden-line emission from CSM knots around the LMC remnants SNR 0519–69.0, N103B, and DEM L71, with electron densities ranging from a few hundred to  $\geqslant$$10^{4}$ cm$^{-3}$ estimated from the [\ion{S}{ii}] doublet. Very recently, \citet{Tenhu2025} identified spatial variations in the electron density of the LMC SNR 0540-69.3 based on the [\ion{S}{ii}] doublet, with the highest densities ($\sim$$10^{4}$ cm$^{-3}$) found in bright knots in the western region and lower densities ($\sim$3$\times$$10^{3}$ cm$^{-3}$) in the eastern part. In addition to the dense-knot CSM scenario, circumstellar structures may also be formed through different mechanisms (see e.g. \citealt{Souropanis2022, Chiotellis2020}). The possible existence of dense CSM around G107.7$-$5.1 and its interaction with the remnant require further observational investigation.

%These knots emit radiation as a result of ionization by multiple shock waves that propagated following the supernova explosion.

%%%%%%%%%%%%%%%%%%%%%%%%%%%%%%%%%%%%%%%%%%%%

The [\ion{O}{iii}] $\lambda$4959 and $\lambda$5007 emission lines are detected in all spectra, and their presence is essential for interpreting the shock structure and constraining the shock velocity based on shock models (e.g. \citealt{CoxRaymond1985, Hartigan1987, Raymond1988}). The measured [\ion{O}{iii}]/H$\beta$ ratios at all slit positions (1.2$-$3.3) indicate that the shocks in G107.7$-$5.1 are associated with complete recombination zones (see \citealt{Raymond1988}).

Shock models further suggest that weak or absent [\ion{O}{iii}] emission is indicative of relatively low-velocity shocks ($\leq80$ km s$^{-1}$). Based on the observed [\ion{O}{iii}]/H$\beta$ ratios across all slit positions and adopting the shock model of \citet{Hartigan1987}, we estimate that the shock velocities are close to 100~km~s$^{-1}$. The highest [\ion{O}{iii}]/H$\beta$ ratio, with a value of $\sim$3.3, is found in the southern region at the S1 slit position, suggesting relatively higher shock velocity compared to the other observed regions. As shown in fig. 1 of \citet{Araya2024}, the gamma-ray emission is more prominent in the southern part of the SNR, which spatially coincides with the [\ion{O}{iii}] emission contours from \citet{Fesen2024}. Although our spectra cover only a slit position (S1) in the southern region, our result is consistent with those reported for the southern region by \citet{Araya2024} and \citet{Fesen2024}.

Model calculations of cooling shocks by \citet{Dopita79} show that the pre-shock density ($n_{0}$) is related to the electron density ($n_{[\ion{S}{ii}]}$) derived from the observed [\ion{S}{ii}] lines by 
\begin{equation}
\label{preshock-density}
n_{[\ion{S}{ii}]} = 45~n_{\rm 0} \times (V_{\rm s}/100~{\rm km~s^{-1}})^{2}~~{\rm cm}^{-3}.
\end{equation}
Assuming a shock velocity of 100~km~s$^{-1}$ and using the derived electron densities, we estimated pre-shock densities in the range of 0.7$-$100 cm$^{-3}$. The higher values within this range are significantly above typical ISM densities, strongly suggesting that the SNR is expanding into regions of locally enhanced ambient density, as observed in other remnants such as G203.1+6.6 \citep{Aktekin2025}, IC~443 \citep{Alarie-Alexandre2019, Bakis2024}, and G190.9$-$2.2 \citep{Bakis2025b}.

The logarithmic extinction coefficient, $c$(H$\beta$), was estimated to fall between 0.1 and 1.6 based on the observed H$\alpha$/H$\beta$ ratios, using the relation  given by \citet{Kaler1976}. Through the conversion $E(B-V) = 0.77c$ \citep{OsterbrockFerland2006}, we obtained reddening values of $E(B-V)$ $\sim$ 0.07$-$1.22. Our spectral data also reveal a significant difference in extinction between the northwestern and southern observed regions, with positions NW2 and S1 exhibiting $A_{\rm V}$ $\sim$ 0.23 and 3.77, respectively. Using the relation of \citet{Predehl1995}, we estimated the hydrogen column density ($N_{\rm H}$) to be (0.4$-$6.6) $\times$ $10^{21}$ cm$^{-2}$, and this range encompasses the total Galactic hydrogen column density along the same line of sight - approximately 4.3 $\times$ $10^{21}$ cm$^{-2}$ - as derived using the \texttt{nhtot}\footnote{\url{https://www.swift.ac.uk/analysis/nhtot/index.php}}
 tool based on the method of \citet{Willingale2013}. Our measurements reveal logarithmic extinction coefficients spanning a broad range of 0.1$-$1.6, resulting in corresponding variations in
 $E(B-V)$, $A_{\rm V}$, and $N_{\rm H}$. Similar wide ranges have been reported for other Galactic SNRs (e.g. G82.2+5.3: \citealt{Mavromatakis2004}, W50: \citealt{Boumis2007}, and HB3: \citealt{Boumis2022}).

%%%%%%%%%%%%%%%%%%%%%%%%%%%%%%%%%%%%%%%%%%%%%%%%%%%%%%%%%%

\subsection{Optical spectral properties of G150.3+4.5}
Similar to G107.7$-$5.1, H$\alpha$, H$\beta$ and the  [\ion{O}{iii}], [\ion{N}{ii}], and [\ion{S}{ii}] doublets are detected at all positions within the southern region of this remnant. The [\ion{O}{i}] doublet is also detected at most of these positions (see Fig.~\ref{fig:figure4}). 

The [\ion{S}{ii}]/H$\alpha$ ratios, ranging from 0.43 to 0.92, together with the [\ion{N}{ii}]/H$\alpha$ ratios of 0.49$-$1.29, indicate that optical emission observed in the southern part of the SNR is consistent with shock-heated gas.

Additionally, the measured ratios from all eight positions of G150.3+4.5 are fully consistent with the diagnostic diagrams \citep{Kopsacheili2020}, confirming that the optical emission originates from shock-heated gas (see Fig.~\ref{fig:figure5}).

The [\ion{S}{ii}] $\lambda$6716/$\lambda$6731 ratio was determined for each slit position and found to span between 0.68$-$1.41 (see Table~\ref{tab:Table3}). At positions S6 and S7, the ratio nearly reaches the theoretical maximum value of $\sim$1.43 \citep{OsterbrockFerland2006}. Based on calculations with the \texttt{PyNeb} code \citep{Luridiana2015}, the inferred electron densities lie between 40 and 2800~cm$^{-3}$ across the S1$-$S8 slit positions. In particular, the electron densities at the S1$-$S4 slit positions exceed 1000 cm$^{-3}$. These high densities are spatially coincident with the southern cloud MC~G150.6+03.7 identified by \citet{Feng2024}, see their fig. 7, suggesting that the elevated electron densities likely arise from shock interaction with dense molecular material. 

The [\ion{O}{iii}]/H$\beta$ ratios measured at all slit positions (1.8$-$4.6) suggest that the shocks propagate through regions with complete recombination zones, corresponding to shock velocities of $\sim$100~km~s$^{-1}$  (e.g. \citealt{Hartigan1987}). At the S1 slit position, where the ratio reaches the upper end of this range ($\sim$4.6), the inferred shock velocity increases to  $\sim$100$-$120~km~s$^{-1}$. This variation suggests the presence of a density gradient in the regions into which the shocks are propagating. Consistent with this interpretation, \citet{Fesen2024} reported substantial differences in the optical emission morphology between the [\ion{O}{iii}] and H$\alpha$ emission lines. Their fig. 19 shows that the southern limb of the SNR is characterized by prominent, sharp [\ion{O}{iii}] filaments and more diffuse emission extending farther south, while no corresponding filamentary H$\alpha$ emission is detected.
Similar morphological differences indicate pronounced inhomogeneities and density fluctuations within the ambient medium, in good agreement with our results (see \citealt{Hester1987, Mavromatakis2002, Boumis2002}). 

Using the equation~(\ref{preshock-density}) and assuming a shock velocity of 100~km~s$^{-1}$, the derived pre-shock densities for G150.3+4.5 ($n_{0}$ $\sim$ 0.9$-$60~cm$^{-3}$) are comparable to those for G107.7$-$5.1, implying a wide range of ambient conditions in both remnants.

The spectra at all slit positions of G150.3+4.5 exhibit considerable variation in extinction, similar to those of G107.7$-$5.1. For example, in G150.3+4.5, while position S2 shows a value of approximately 1.6, the extinction at position S7 reaches 3.5 (see Table~\ref{tab:Table3}). The extinction variations observed in both G150.3+4.5 and G107.7$-$5.1 suggest the presence of substantial dust structures along this line of sight.

Our optical spectroscopic analyses reveal that these two remnants exhibit notable similarities in several diagnostic spectral properties, including the physical conditions of both the pre-shock and post-shock gas as well as their shock characteristics. Their GeV gamma-ray properties were also reported to be comparable by \citet{Araya2024}. Both SNRs are radio-dim and located at relatively high Galactic latitude, which may suggest that they are evolving in low-density environments \citep{Araya2024}. On the other hand, the southern region of G150.3+4.5 is particularly interesting, as it exhibits an association with MCs \citep{Feng2024} and signatures of hadronic gamma-ray emission \citep{Li2024}. Despite the need for more detailed observations of these large-angular-size remnants and their environments, their known properties make them valuable examples of evolved SNRs detected at GeV energies but lacking prominent radio counterparts (see table 4 of \citealt{Burger-Scheidlin2024}).

%%%%%%%%%%%%%%%%%%%%%%%%%%%%%%%%%%%%%%%%%%%%%%%%%%%%%%%%%%%%%%%%%%%%%%%%%%%%
\section{Conclusions}
\label{sec:conc}
We present the results of the first detailed optical spectroscopic investigation of the SNRs G107.7$-$5.1 and G150.3+4.5, from which we draw the following conclusions:

\begin{enumerate}

\item The SNR nature of G107.7$-$5.1 and G150.3+4.5 is supported by elevated [\ion{S}{ii}]/H$\alpha$ (0.56$-$0.86 and 0.43$-$0.92) and [\ion{N}{ii}]/H$\alpha$ (0.7$-$1.4 and 0.49$-$1.29) ratios,
which are characteristic of shock-heated gas. 
These emission-line ratios, together with other observed line ratios such as [\ion{O}{i}] $\lambda$6300/H$\alpha$ and [\ion{O}{iii}] $\lambda$5007/H$\beta$, are fully consistent with diagnostic diagrams that distinguish shock excitation from photoionization, providing compelling evidence that the optical emission in both remnants is dominated by shock-heated gas.

\item Analysis of the [\ion{S}{ii}] $\lambda$6716 and $\lambda$6731 lines in all spectra indicates that the electron density ($n_{\rm e}$) in G107.7$-$5.1 ranges approximately from 30 to 4650 cm$^{-3}$, while in G150.3+4.5 it ranges from roughly 40 to 2800 cm$^{-3}$, showing significant variations across the observed regions.

\item  The observed [\ion{O}{iii}]/H$\beta$ ratios for both SNRs indicate the presence of shocks with complete recombination zones and are consistent with shock velocities of $\sim$100~km~s$^{-1}$ or higher. 

\item The extinction variations also observed in the optical spectra of both SNRs may indicate the presence of substantial dust structures along the line of sight.

\end{enumerate}

We conclude that these two remnants display remarkable similarities across multiple optical diagnostic spectral properties. Future detailed observations of G107.7$-$5.1, particularly in the radio and X-ray regimes, are needed to better constrain its physical properties  - including its distance and progenitor characteristics -  and to further advance our understanding of SNR evolution. High-resolution spectroscopy of G107.7$-$5.1 and G150.3+4.5 with {\it XRISM} \citep{Tashiro2025}, as well as future observatories such as {\it NewAthena} \citep{Cruise2025} and {\it AXIS} \citep{Reynolds2023}, will be crucial for accurately characterizing the emission processes of these SNRs.

%%%%%%%%%%%%%%%%%%%%%%%%%%%%%%%%%%%%%%%%%%%%%%%%%%%%%%%%%%%%%%%%%%%%%%%%%%%%
\section*{Acknowledgements}
We thank the referee, Panayotis Boumis, for the valuable comments and insightful suggestions, which have significantly improved the quality and clarity of this manuscript. We also thank Robert Fesen for providing the optical images of G107.7$-$5.1 and G150.3+4.5. This research was supported by the Scientific and Technological Research Council of T\"{u}rkiye (T\"{U}B\.{I}TAK) through project number 124F089. In this study, observational data obtained within the scope of the project numbered 25ARTT150-3015, conducted using the RTT150 Telescope and the TFOSC system at the TUG (T\"{U}B\.{I}TAK National Observatory, Antalya) site under the T\"{u}rkiye National Observatories, have been utilized, and we express our gratitude for the invaluable support provided by the T\"{u}rkiye National Observatories and their personnel. 

%%%%%%%%%%%%%%%%%%%%%%%%%%%%%%%%%%%%%%%%%%%%%%%%%%
\section*{Data Availability}
The optical data from the RTT150 telescope will be shared on reasonable request to the corresponding author.



\bibliographystyle{mnras}
\bibliography{example} % if your bibtex file is called example.bib

%\begin{thebibliography}{99}
%\end{thebibliography}

%%%%%%%%%%%%%%%%%%%%%%%%%%%%%%%%%%%%%%%%%%%%%%%%%%

% Don't change these lines
\bsp	% typesetting comment
\label{lastpage}
\end{document}